# DEPLETION-DRIVEN MORPHOLOGICAL TRANSITIONS

# IN HEXAGONAL CRYSTALLITES OF VIRUS RODS

B. Sung,[1] H. H. Wensink,[2] E. Grelet,[1,*]

[1] *Centre de Recherche Paul Pascal, UMR 5031,*

*CNRS & Université de Bordeaux, 33600 Pessac, France*

[2] *Laboratoire de Physique des Solides, UMR 8502,*

*CNRS & Université Paris-Sud, Université Paris-Saclay, 91405 Orsay, France*

**Abstract**

The assembly of nanometer-sized building blocks into complex morphologies is not only of fundamental interest but also plays a key role in material science and nanotechnology. We show that the shape of self-assembled superstructures formed by rod-shaped viruses can be controlled by tuning the attraction via the depletion interaction between the rods. Using non-adsorbing polymers as a depleting agent, we demonstrate that a hierarchical unidimensional self-organization into crystalline clusters emerges progressively upon increasing depletion attraction and enhanced growth kinetics. We observe a polymorphic change proceeding from two-dimensional (2D) crystalline monolayers at weak depletion to one-dimensional (1D) columnar fibers at strong depletion, via the formation of smectic fibrils at intermediate depletion strength. A simple theory for reversible polymerization enables us to determine the typical bond energy between monomeric units making up the smectic fibrils. We also demonstrate that gentle flow-assistance can be used to template filament-like structures into highly aligned supported films. Our results showcase a generic bottom-up approach for tuning the morphology of crystalline superstructures through modification of the interaction between non-spherical building blocks. This provides a convenient pathway for controlling self-organization, dimensionality and structure-formation of anisotropic nanoparticles for use in nanotechnology and functional materials.

**Introduction**

The ability to control the shape and symmetry of structures formed through self-assembly processes is of fundamental importance in fabricating highly-ordered nano- and micro-structures (1) (2). The large diversity of morphologies emerging from such superstructures finds applications in photonics (3) (4) (5), electronics (6) (7), sustainable energy (8) (9) (10), drug delivery (11) (12) and bioengineering (13) (14). Finding ways to guide nanoparticle self-assembly into mono- and multi-layer structures, bundles, tubes, spherical or polyhedral shells, and more complex geometries has been one of the most pursued research themes in nanoscience. One of the main driving forces for studying controlled nanoparticle self-assembly is the many applications in nanotechnology ranging from tunable optical properties (15) (16), anisotropic electron transport (17) (18), to functional surface topography (19) (20). Among such complex architectures, unidimensional (1D) fibers and bidimensional (2D) superlattices composed of rod-shaped building blocks are of particular interest in view of their intrinsic anisotropic properties. For example, fibers of 1D aligned carbon nanotubes, obtained from various spinning techniques, exhibit high mechanical stiffness and outstanding electrical performance (21) (22) (23). Moreover, 2D long-ranged crystalline monolayers of inorganic nanorods can be used for generating surface plasmonics or piezoelectricity, or for creating photonic crystals (24) (25) (26) (27).

Contrary to spherical nanoparticles, reports on controlled self-assembly of rod-shaped particles are scarce, and no clear indications of polymorphisms in their self-assembled structures have been adequately demonstrated so far. In view of this, we address the following two questions. First, what variety of ordered anisotropic superstructures can be expected if we were able to systematically tune the interparticle forces in a colloidal suspension of rods? And more specifically, is it possible to define a pathway enabling us to control the shape of these self-assembled objects from predominantly unidimensional (fiber-like) to bidimensional (disk-shaped) without compromising their long-range crystalline order? For this purpose, the depletion interaction (31) appears to be a suitable and versatile mechanism for controlling particle interaction, as the range and strength of depletion interaction are determined by the polymer size and concentration, respectively (32). We will show that introducing tunable depletion interactions with a suitable range indeed leads to a high degree of controlled self-assembly generating a wide variety of crystallite morphologies without the need to use microscale confinements (28) or interfacial effects (29) (30).

It is well known that adding non-adsorbing polymers to colloidal suspension provides an efficient method to induce effective particle-particle attraction that can direct hierarchical organization processes (33) (34) (35) (36) (37) (38) (39). Experimental and theoretical studies on model systems of rod-like particles such as fd viruses have established that the experimental conditions are such that the formation of partially-ordered structures, such as nematic tactoids, membranes and their stacks, as well as twisted ribbons (37) (40) are strongly favored. These depletion-driven self-assembled morphologies have been obtained using high molecular-weight depletants such as Dextran (molecular weight, $M_w$ = 500,000 g/mol; radius of gyration, $R_g$ ~21 nm), whose size is larger than the particle diameter. This results in the virus rods self-assembling into super-structures with predominant liquid-like or short-ranged positional order in at least two spatial directions (37) (40) (41) (42). However, in order to fully benefit from the morphology effect in terms of enhanced anisotropic properties of the resulting materials, highly ordered structures with crystalline, i.e. long-ranged positional order are required (4) (5) (10).

In order to promote dense self-assembled structures with crystalline order short-ranged depletion attractions are required favoring small inter-rod distances, s. This effect can be achieved using a non-adsorbing polymer of small molecular weight (PEG, $M_w$ = 8,000 g/mol; $R_g$ ~4 nm) added to a suspension of fd virus. The key-point here is that corresponding polymer coil has about the same size as the virus diameter, $2.R_g \sim d$ =7 nm. Taking advantage of the micrometer length of fd virus we are able to analyze the structures formed by these filamentous rods at the single particle scale using fluorescence optical microscopy (39) (43). We then investigate the detailed morphology of the virus-based superstructures upon varying both the depletion strength (polymer concentration) and rod volume fraction, while keeping the polymer size fixed thus ensuring the rod-rod interactions to be short-ranged.

**Results and discussion**

The experimental results are presented in Figure 1. Here, we demonstrate that subtle morphological changes occur in a hierarchical way upon increasing the depletion strength. At low polymer concentration, when the depletion attraction is weak, the rods tend to aggregate side-by-side (44) thus minimizing their excluded volume. This results in the formation of 2D plate-shaped crystallites. These platelets have well-developed hexagonal facets (Figure 1a) and are composed of a single layer of hexagonally packed aligned viral rods with a protruding needle-like defect located at the core (43). The core-defect acts as a steric deterrent preventing the platelets to stack on top of each other.

Enhancing the depletion attraction causes an increase in the density number $\rho_0$ of plate-shaped nuclei and a simultaneous decrease of the platelet diameter (D=1-2 μm) as well as a length reduction of the central core defect (Figures 1b and S2). This results in a decrease of the average platelet diameter D with increasing the depletion strength (Figure S3), since $\rho_0 \propto 1/D^2$ as dictated by particle mass conservation.

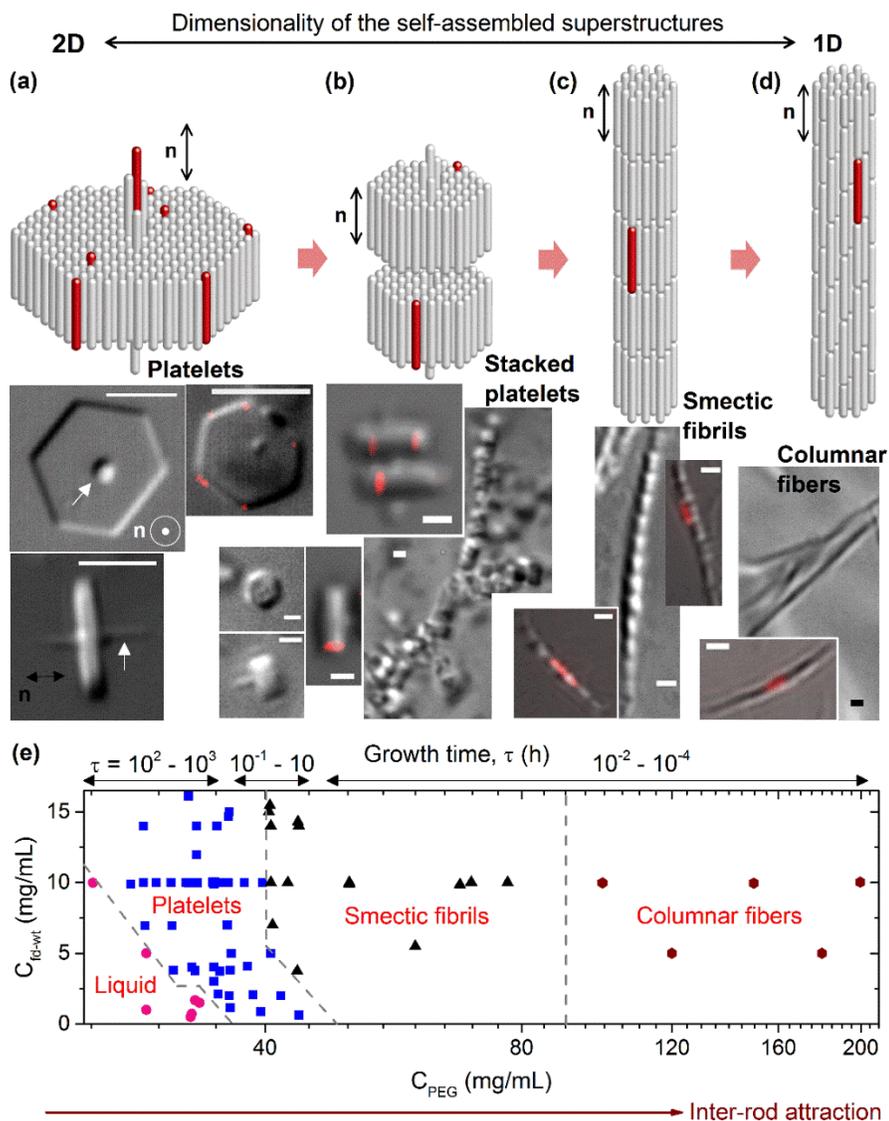

**Figure 1. Overview of morphological and dimensional changes in self-assembled structures formed by rod-like fd viruses driven by increasing depletion attraction**. The kinetic processes determining the typical growth time for the various structures speeds up considerably with increasing inter-rod attraction. The self-assembled morphology changes from (a) hexagonal platelets composed of a one-rod thick monolayer of aligned fd rods with a central core defect (indicated by a white arrow), (b) small hexagonal platelets and their stacks. When the attraction increases further, the rods self-assemble into (c) smectic fibrils composed of small polymerized platelet monomers, and then to (d) columnar fibers. The fluorescently labeled viruses in the DIC/fluorescence overlaid images show that the rods are aligned parallel to the main axis **n** of the cluster (black double arrow). Scale bars, 5 µm for (a) and 1 µm for (b-d). (e) Phase diagram as a function of the rod ($C_{fd-wt}$) and polymer ($C_{PEG}$) concentrations, where the typical growth time $\tau$ is also indicated. At each phase boundary, a gradual change of morphology occurs rather than a sharp transition.

Upon increasing the depletion strength, the self-assembly process speeds up considerably resulting in much shorter typical growth times. Pushing the depletion strength even further we observe a high population of small platelets (of about D~1 μm). Without a detectable central defect preventing their clustering, the small platelets begin to stack up and finally self-assemble into smectic fibrils in which each layer represents a small hexagonal platelet with a width of about one rod length (Figure 1c). This quasi-1D morphology referring to the overall resulting shape of the self-assembly can be interpreted as the result of fast-growing hexagonal platelets reversible polymerizing into smectic filaments. Clearly, the effective depletion attraction acting between platelets promotes face-to-face stacking in order to maximize the free volume available to the polymer (32). In order to estimate the typical bond energy between monomeric platelets making up a smectic fibril, we invoke a simple reversible polymerization model described in the Supporting Information. This theory assumes the smectic fibrils to operate in thermodynamic equilibrium in which case the fibril height distribution can be determined analytically and compared to experimental results depicted in Fig S4a. From the fits (see Supporting Information) we estimate a typical bonding energy between monomeric platelets of about 10 $k_BT$ and a typical 'bonding volume' of effective range of attraction of about 8-10 times the polymer volume. The 'bonding volume' which can be interpreted as the bonding range times the platelet area is considerably less than the volume ~ $R^2.R_g$ with R the radius of a flat plate one would naively obtain from free-volume theory (32) if the face-to-face bonding was entirely due to ideal depletion attractions. Complicating factors such as thermal corrugation of the plate surface (due to fluctuations of the rods around their centers of mass), electrostatic interactions and semi-flexibility of the filamentous virus rods may account for the strongly reduced bond energy but these effects are notoriously difficult to capture within a simple, tractable model.

At the highest depletion attraction the growth kinetics speeds up dramatically resulting in the rods self-assembling into columnar bundles (Figure 1d). In these structures, there is no layered organization along the main axis of the fiber but the hexagonal crystalline structure transverse to the bundle director **n** is preserved. The time required for the self-assembly processes, or growth time, is reduced dramatically when the growth morphology changes from the predominantly bidimensional platelets to the unidimensional fibers and bundles.

Throughout the range of cluster morphologies observed, the typical growth time varies by more than six orders of magnitude, taking up days for large platelets to form, a few minutes for the smectic fibrils, down to milliseconds for the fast formation of columnar fibers. Apart from the intrinsic rod-rod attraction, kinetics play an important role in templating the self-assembled structures obtained. In view of the ultrafast growth process, we expect the formation of columnar bundles to be strongly marked by non-equilibrium effects, as columnar structures composed of long rods usually do not emerge from thermodynamic considerations (32) (46). In all the observed superstructures, the orientation of the rods (**n**) remains parallel to the main axis of the crystallite, as evidenced from the presence of a small fraction of virus rods labelled with a red fluorescent dye (Figure 1) and using a full-wavelength retardation plate (Figure S1). The absence of any detectable particle self-diffusion within the clusters (Movies S1 to S5) is consistent with the long-range positional order along the direction normal to **n** observed by small-angle X-ray scattering (SAXS) experiments (Figure 2).

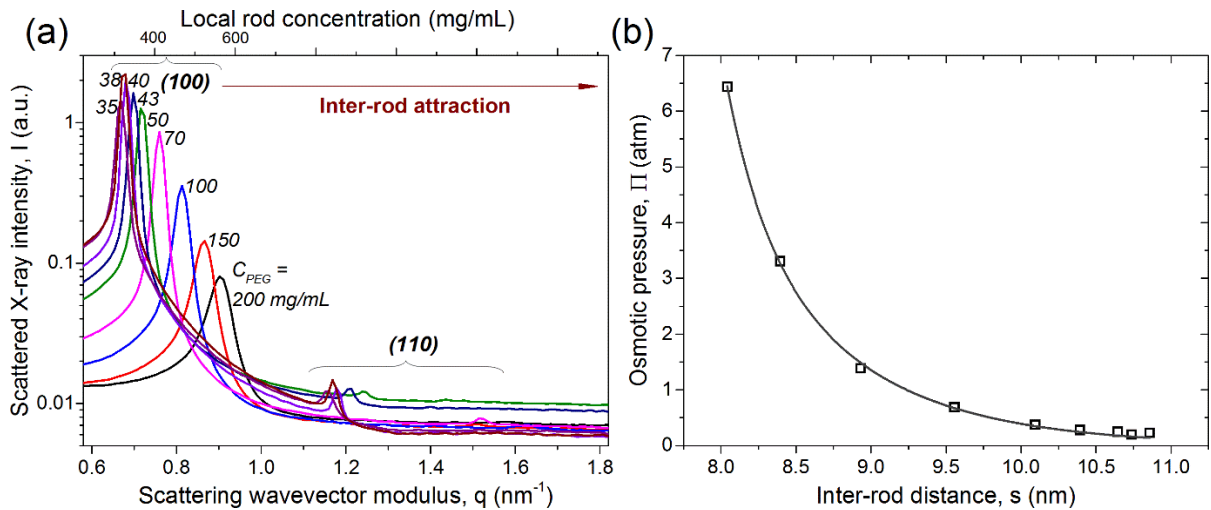

**Figure 2. SAXS data of the different virus-based self-assembled superstructures obtained by varying the concentration of PEG depletant ($C_{PEG}$).** The initial virus concentration is set at 10 mg/mL. (a) SAXS spectra characteristic of hexagonal long-ranged positional order with the presence of sharp 100 and 110 Bragg reflections. The virus concentration *within* the self-assembled clusters, referred to as the local concentration, is calculated from the 2D swelling law associated with a hexagonal ordering of long rods (46) for a given (100) peak position. (b) Osmotic pressure ($\Pi$) obtained from the PEG concentration (47) as a function of inter-rod distance, $s = (4\pi/\sqrt{3})q_{100}^{-1}$. The fit (black curve) is based on the model of Yasar et al. (48) described in the Supporting Information.

The SAXS experiments have been conducted at a fixed rod concentration but at varying polymer concentration and depletion strength. All SAXS spectra (Figure 2a) display (100) and (110) diffraction peaks, characteristic of local hexagonal rod packing. The interaxial distance between the neighboring rods $s = (4\pi/\sqrt{3})q_{100}^{-1}$ can be directly obtained from the position of the main diffraction reflection. There is only little variation of the distance s (Figure 2b) accompanying the vast morphological changes observed throughout the same range of conditions. Since the interaxial rod distance s probed by SAXS is about two orders of magnitude smaller compared to the rod length $\ell$, a 2D swelling law ($q_{100} = \eta \cdot C_{rod}^{1/2}$) of the hexagonal lattice is expected, with $\eta = \left(8\pi^2 N_A \ell /(\sqrt{3}\, Mw)\right)^{0.5}$ (46). From this, the local rod concentration associated with the various self-assembled morphologies can be deduced. This gives a value of around 300 mg/mL for the platelets, 350-400 mg/mL for the smectic fibrils, and 450-550 mg/mL for the columnar bundles (Figure 2a).

The depletion interaction between the virus rods can be accounted for by invoking a polymer-generated osmotic pressure $\Pi$ acting on the virus arrays. Here, one simply assumes that most of the polymer is excluded from the virus-based superstructures (45). Such models, described in the Supporting Information, have been successfully applied to many charged rod-like particles (46) (47), and give an accurate description of the experimental data as shown in Figure 2b.

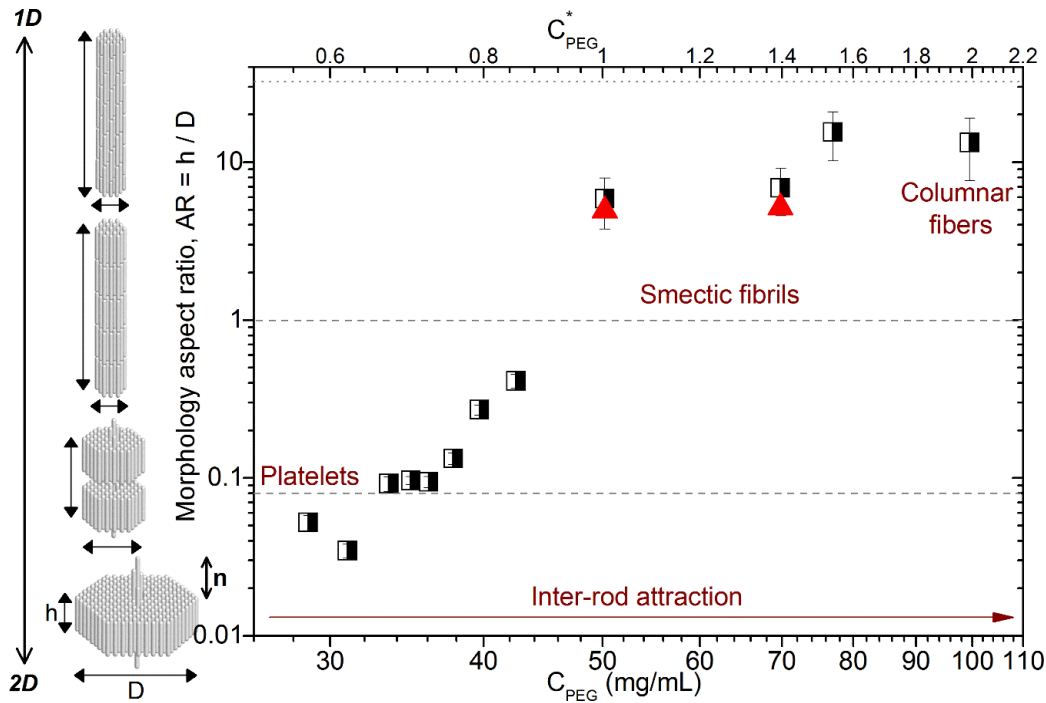

**Figure 3. Morphological transition from 2D to 1D self-assembled structures observed by enhancing the inter-rod attraction, in turn controlled by the strength of the depletion interaction.** The cluster aspect ratio AR, defined as the dimension parallel to **n** (h; height) over the dimension perpendicular to **n** (D; diameter) where **n** denotes the main axis of the cluster morphology, is plotted (black and white squares) as a function of the depletant concentration. Experimentally, the largest measurable length is around 20 µm (the corresponding aspect ratio is indicated by a dotted line), since longer fibrils no longer lie within the same focal plan. The red triangular symbols represent the theoretical predictions from a simple reversible polymerization theory describing the formation of smectic fibrils (see Supporting Information), and they show a good quantitative agreement with experimental data. The typical bond energy between monomeric platelets within the smectic fibrils is estimated to be about 10 $k_B T$.

Figure 3 displays the crossover from one self-assembled superstructure to another in terms of the aspect ratio AR=h/D of the typical cluster morphology, with h and D indicating the height and diameter of the superstructure measured parallel and perpendicular to the rod director **n**, respectively. Throughout the range of attractions, the diameter decreases from 20-30 µm for the platelets while reaching a minimum of around 0.9 µm for the smectic fibers and columnar bundles (Supporting Information, Figures S3 and S4). The height increases from one rod-length for single platelets to 10-20 µm for smectic fibers and columnar bundles (Supporting Information, Figures S3 and S4). Consequently, the aspect ratio AR of the cluster morphology varies by almost three orders of magnitude throughout the probed range of depletion strengths. This demonstrates a high degree of shape-tunability of the self-assembled clusters, while guaranteeing the long-ranged positional order within the structures as demonstrated by the SAXS experiments.

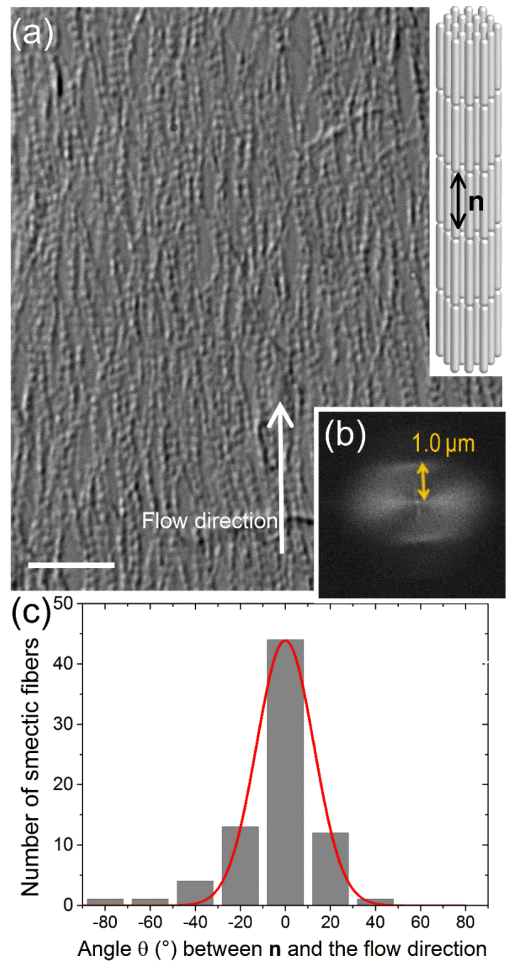

**Figure 4. Flow assisted alignment of smectic fibrils deposited on a glass substrate.** (a) DIC image of the smectic fibrils, whose main axes (**n**) are along the flow direction. Scale bar, 10 µm. (b) Fourier transform pattern of the DIC image in (a), with the one-rod-thick smectic layer spacing indicated. (c) Distribution of smectic fibril orientations fitted by a Gaussian function (red line), from which the 2D

orientational order parameter, $S=\langle 2.\cos^2(\theta)-1\rangle=0.91$, can be obtained after numerical integration.

Films of highly aligned smectic fibers can be obtained upon applying weak shear flow, as shown in Figures 4a. The degree of alignment can be established from a 2D nematic order parameter $S=\langle 2.\cos^2(\theta)-1\rangle$ that we can extract from the orientational distribution of smectic fibrils (Figure 4c). Typical values we found are of about S=0.90 (note that S=1 indicates perfect alignment, and S=0 complete orientational disorder). This shows that these self-assembled superstructures are strongly susceptible to external orientation stimuli and can be readily shaped into macroscopically aligned domains. The facile processability of these rod-based superstructures opens perspectives for applications requiring anisotropic stimuli-responsive materials with internal crystalline order (9) (10) (14) (29) (50).

**Conclusion**

In conclusion, we have described a tunable self-assembly process of rod-like particles forming superstructures with an unprecedented control of their cluster morphology, effective dimension and internal microstructure. In our system of filamentous fd virus rods mixed with PEG polymer, controlled self-assembly is achieved through the use of tunable short-ranged depletion interactions between the rods generated by the presence of non-adsorbing polymer with typical size comparable to the rod diameter. A systematic increase in the polymer concentration causes a number of morphological transitions affecting the principal dimensionality of the self-assembled structure. These range from isolated quasi-bidimensional platelets and stacked oligomeric platelets at weak depletion to unidimensional polymeric smectic fibrils, and columnar fibers at strong depletion. These polymorphological changes occur without compromising the long-ranged hexagonal crystal order within the superstructures. The average inter-rod distance is only weakly affected by the depletion strength. Simple thermodynamic considerations enable us to predict the typical fibril length and infer the typical bond energy between monomeric platelets within these structures. Last but not least, we have demonstrated the possibility of flow-assisted templating of such anisotropic superstructures into highly aligned supported films demonstrating a high degree of processability. By expanding the spectrum of bottom-up approaches towards anisotropic nanoparticles we argue that depletion-driven rod suspensions displaying hierarchically-tunable morphogenesis constitute a promising candidate for fabricating self-assembled structures for use in devices depending on anisotropic stimuli-responsive materials.

**Materials and Methods**

*Virus stock suspensions*

Wild-type fd bacteriophages were grown with ER2738 strain of *E. coli* as host bacteria, and were purified according to standard protocols (51). The fd-wt (wild type) viruses are rod-like particles, monodisperse in size and shape, with contour length $\ell$ = 880 nm and diameter $d$ = 7 nm. In suspension, the virus rods have been shown to behave as near-hard rods exhibiting a well-defined liquid-crystalline phase behavior (41) (39) (40) (46). Virus suspensions were extensively dialyzed against TRIS-NaCl-HCl buffer (pH 8.2) at 110 mM of ionic strength, to ensure the electrostatic repulsion between viral particles is strongly screened (Debye screening length, $\lambda_D$ = 0.9 nm). The rods were subsequently concentrated using ultracentrifugation and redispersed at 30-35 mg/mL in the same buffer solution. The virus concentration was determined using spectrophotometry at the peak absorption wavelength of 269 nm with an optical density of 3.84 cm$^2$/mg (52). Fluorescently-labeled virus batches were separately prepared by conjugating Alexa Fluor 488-TFP ester (Invitrogen) or Dylight550-NHS ester (ThermoFischer) to their coat proteins. These labeled viruses were added in tracer amounts (0.001% (w/w) to 0.1% (w/w)) to the non-labeled virus batches for tracking by fluorescence microscopy.

*Sample preparation*

Polyethylene glycol (PEG) of $M_w$ = 8,000 g/mol (Sigma-Aldrich) was used as a non-absorbing polymer with a radius of gyration ($R_g$) of 4 nm in aqueous solution (53). All polymer-virus mixtures were prepared in TRIS-NaCl-HCl buffer (pH 8.2) adjusted at 110 mM of ionic strength, and injected into optical microscopy cells. The cells were made by a glass slide and a coverslip (initially cleaned with sulfochromic acid) separated by a Mylar or Parafilm spacer, to obtain a cell thickness of about 100 μm. The cells were then sealed with UV-cured glue (NOA81, Epotecny). For the film of aligned smectic fibers, the samples were either prepared by drop casting on a cover slip, or loaded and oriented in the cell through capillary forces. Samples for small-angle X-ray scattering (SAXS) experiments were prepared in cylindrical quartz capillaries (diameter ~1.5 mm; Mark-Röhrchen), filled with aqueous virus-polymer mixtures (~20 μl for each), and sealed by flame. The capillaries were positioned vertically (gravity along the main axis of a capillary) for about 4 weeks in order to induce macroscopic phase separation through sedimentation of the self-assembled structures at the capillary bottom.

*Optical microscopy*

Differential interference contrast (DIC) and epifluorescence images were obtained using an inverted microscope (IX71, Olympus) equipped with an oil-immersion objective (NA 1.4, ×100

UPLSAPO), a mercury-halide excitation source (X-cite120Q, Excelitas), and a fluorescence imaging camera (NEO sCMOS, Andor Technology). To enable simultaneous acquisition of DIC and fluorescence images, an optical splitter setup (Optosplit II, Cairn Research) was used to divide each image into two channels thanks to a dichroic beamsplitter and band pass filters. The images obtained from both channels were then overlaid. The whole imaging system was operated by computer-interface software (Meta-Morph, Molecular Devices).

*Small angle X-ray scattering (SAXS)*

SAXS measurements were performed at the SWING beamline at the synchrotron facility SOLEIL (Orsay, France) operating at a source wavelength of $\lambda = 0.0995$ nm. The diffraction pattern was recorded by an AVIEX CCD detector, which was located in a vacuum detection tunnel with a sample-to-detector distance of 1.49 m. Angular integration was applied on the 2D SAXS pattern to obtain the scattering intensity as a function of scattering vector modulus, q.

**Acknowledgments:** This work was supported by the French National Research Agency (ANR) through the project AURORE. We thank J. Perez from the SWING beamline for his kind assistance.

**Author contributions:** B.S. prepared the samples, and performed the experiments. B.S. and E.G. analyzed the data. H.H.W. developed the theoretical models. E.G. conceived of the project, designed experiments and fit the data. All authors wrote the manuscript.

**Conflicts of interests:** There are no conflicts to declare.

**Corresponding author:** eric.grelet@crpp.cnrs.fr